\documentclass[aps,prl,floatfix,epsfig,twocolumn,showpacs,preprintnumbers]{revtex4}
\usepackage{amssymb}
\usepackage{graphicx}
\usepackage{amsmath}

\DeclareMathAlphabet      {\mathbf}{OT1}{cmr}{bx}{n}
\begin{document}

\title{Local Self--Energies for V and Pd Emergent from a Non--Local LDA+FLEX
Implementation}
\author{Sergey Y. Savrasov, Giacomo Resta}
\affiliation{Department of Physics, University of California, Davis, CA 95616, USA}
\author{Xiangang Wan}
\affiliation{National Laboratory of Solid State Microstructures and Department of
Physics, Nanjing University, Nanjing 210093, China}
\affiliation{Collaborative Innovation Center of Advanced Microstructures, Nanjing
University, Nanjing 210093, China}

\begin{abstract}
In the spirit of recently developed LDA+U and LDA+DMFT methods we implement
a combination of density functional theory in its local density
approximation (LDA) with a $k$-- and $\omega -$dependent self--energy found
from diagrammatic fluctuational exchange (FLEX) approximation. The active
Hilbert space here is described by the correlated subset of electrons which
allows to tremendously reduce the sizes of matrices needed to represent
charge and spin susceptibilities. The method is perturbative in nature but
accounts for both bubble and ladder diagrams and accumulates the physics of
momentum resolved spin fluctuations missing in such popular approach as GW.
As an application, we study correlation effects on band structures in V and
Pd. The d--electron self--energies emergent from this calculation are found
to be remarkably k--independent. However, when we compare our calculated
electronic mass enhancements against LDA+DMFT, we find that for a long
standing problem of spin fluctuations in Pd, LDA+FLEX delivers a better
agreement with experiment, although this conclusion depends on a particular
value of Hubbard $U$ used in the simulation. We also discuss outcomes of a
recently proposed combinations of k--dependent FLEX with DMFT.
\end{abstract}

\maketitle

\section{I. Introduction}

Although electronic structure calculations utilizing a combination of local
density functional and dynamical mean field theories (so called LDA+DMFT)
became a method of choice for studying realistic correlated electrons
systems \cite{RMP}, the search for a more accurate treatment of the electron
self--energy continues to be an active field in the many--particle physics
of condensed matter. Since DMFT accounts for local electronic correlations
by corresponding solution of the Anderson impurity problem, extensions of
this method to small clusters, such as Dynamical Cluster Approximation (DCA)%
\cite{DCA} or cellular version of DMFT (C--DMFT) \cite{CDMFT} have been
proposed in the past and more recently, two promising developments, a dual
fermion (DF) approach \cite{DF} and a dynamical vertex approximation (D$%
\Gamma $A) \cite{DGA}, have been elaborated. Unfortunately, combinations of
these approaches with LDA and their applications to real materials are
complex, time consuming and have so far been scarce \cite{CDMFT-APPS}.

One recent development is the combination \cite{GW+DMFT} of the DMFT with
much celebrated GW approach \cite{GW} that, in contrast to\ LDA+DMFT, tries
to treat dynamically screened Coulomb interaction from first principles
using a summation of a series of the particle--hole bubble diagrams, and
does not rely on any \textit{ad hoc} parametrizations such as "Hubbard U".
Another advantage of the method is the treatment of the local self--energy
using DMFT\ and the access to its k--dependence via the use of the GW
diagram. Interestingly however, that the GW itself misses an important
physics associated with a well--known electronic mass renormalization due to
paramagnons which has been recovered in the summation of the so called
particle--hole ladder diagrams long time ago \cite{Paramagnons}. Thus,
including both the bubble and ladder diagrams in a unified way with DMFT\
may provide more accurate interpolation for the k--dependence of the
self--energy.

There was indeed a very long interest in such development. The paramagnons
or spin fluctuations (SF) were originally shown to suppress electron--phonon
pairing in conventional superconductors \cite{BerkSchrieffer} but capable of
producing d--wave pairing in unconventional superconductors \cite%
{Scalapino,Varma}. The proposed Fluctuational Exchange Approximation (FLEX) 
\cite{FLEX} includes both particle--hole ladders and bubbles as well as
particle--particle ladder diagrams, with the latter contribution being found
of lesser importance at least in some models \cite{Muller}. It was the
method of choice in many studies of strongly correlated systems in the past%
\cite{FLEX-Review}. A proposal to combine FLEX with LDA in a form of a LDA++
method came out in the earlier days of the LDA+DMFT\ developments \cite%
{LDA++}. Very recently, a combination of FLEX and DMFT on the level of a
single--band Hubbard model in 2D was shown to reproduce a dome--like
behavior of critical temperatures characteristic of cuprates superconductors 
\cite{FLEX+DMFT}.

Unfortunately, a completely first principles treatment of the ladder
diagrams represents a challenge. First, the response functions in such
theory are all four--point functions in real space and not two--point
functions as in the GW method. Second, and most important, it has long been
recognized that it is not the bare Coulomb but some semi--phenomenological
Stoner--like interaction $I$ should represent the short--range repulsion\
between correlated electrons in the ladder series\cite{Paramagnons}. The
access to this quantity within local density functional theory has allowed
to compute spin fluctuational spectra together with paramagnon enhanced
self--energies in V and\ Pd \cite{Winter}. A more recent formulation of the
method using all--electron basis sets and accurate summations over
unoccupied states \cite{Sternheimer} has also allowed a few applications to
\ selected transition metals \cite{SpinFluct}.

Despite earlier excitement, the LDA based approaches to spin fluctuations
did not receive a wider recognition because being a homogeneous electron gas
theory LDA may seriously underestimate $I$ for many correlated systems, and
it has been stressed that the Hubbard parameter $U$ should be used instead%
\cite{LDA+U}. This gave rise to a famous LDA+U method\cite{LDA+U-Review}.

At the absence of a first principles treatment of the interaction that
enters the ladder diagrams we still have to rely on parametrizations in
terms of $U$. However, the remaining part of the algorithm is implementable
in principle: the FLEX\ self--energies for correlated electrons, $\Sigma
^{FLEX}(\mathbf{k},\omega )$ can be computed using contributions from both
bubble and ladder diagrams, and subsequently combined with LDA using a
method of projectors. This is exactly as it is done in LDA+U\ and LDA+DMFT.
Furthermore, the method can be extended by adding DMFT, as it has been
recently shown for models \cite{FLEX+DMFT,Danger}. The k--dependent
corrections within such method are attributed to the diagrammatic FLEX which
captures the important physics of spin fluctuations and is simpler to
implement than recently proposed DF and D$\Gamma $A approaches. As the
danger of partial diagrammatic summations was pointed out in a recent work%
\cite{Danger} we bear in mind that such schemes are perturbative in spirit
and should not be literally used for any $U$.

In this work we describe an implementation of such method and provide
applications to V and Pd. It is quite remarkable that the non--local FLEX
self--energies that we extract in our implementation are fairly
k--independent which justifies the use of local approximations. We calculate
the mass enhancements of the d--electrons and compare them against LDA+DMFT
calculation performed with numerically exact Continuos Time Quantum Monte
Carlo (QMC) method\cite{CTQMC} and other published DMFT calculations \cite%
{PdDMFT}. We find that FLEX delivers larger electronic masses than DMFT and
agrees better with experiment, however, this conclusion depends on the value
of $U$ that is used in the simulation. We also comment on a performance of
recently proposed DMFT+FLEX schemes \cite{FLEX+DMFT,Danger} to the problem
of mass enhancement in these two metals.

The paper is organized as follows. In Section II,\ we provide a general
description how the self--energy for correlated electrons is combined with
LDA\ (SELDA\ family of methods) and give specific details about our FLEX
implementation. Various forms of self--consistency conditions are also
discussed. In Section III,\ our applications for V and Pd are described. In
Section IV, we conclude with the perspective on possible applications of
such method to other systems.

\section{II. Family of SELDA\ Methods}

The family of approaches that combine the self--energy for correlated
electrons with LDA\ (the SELDA\ family), relies on a separation of sites
given by the positions $\{\mathbf{\tau }\}$ inside the unit cell of the
lattice onto uncorrelated and correlated sites denoted hereafter by the
positions $\{\mathbf{t}\}$. The site dependent projector operators are
introduced with help of radial solutions $\phi _{at}(\mathbf{r}_{t})=\phi
_{lt}(r_{t})i^{l}Y_{lm}(\hat{r}_{t})$ (where $\mathbf{r}_{t}=\mathbf{r}-%
\mathbf{t})$ of the one--electron Schroedinger equation taken with a
spherically symmetric part of the full potential. \cite{phidot}. The Hibert
space \{$a$\} inside the designated correlated site $\mathbf{t}$ may further
restrict the full orbital set by a subset of correlated orbitals, such,
e.g., as 5 for $l=2$ or 7 for $l=3$. Here, we keep the spin index
explicitly, therefore treat the non--local self--energy $\hat{\Sigma}%
_{\sigma \sigma ^{\prime }}\equiv \Sigma _{\sigma \sigma ^{\prime }}(\mathbf{%
r},\mathbf{r}^{\prime },\omega )$ as the matrix in spin space. It is viewed
in travelling wave representation 
\begin{equation}
\Sigma _{\sigma \sigma ^{\prime }}^{\mathbf{k}}(\mathbf{r}_{\tau },\mathbf{r}%
_{\tau ^{\prime }}^{\prime },\omega )=\sum_{R}e^{i\mathbf{kR}}\Sigma
_{\sigma \sigma ^{\prime }}(\mathbf{r},\mathbf{r}^{\prime }-\mathbf{R}%
,\omega )  \label{EQ21}
\end{equation}%
as follows

\begin{widetext}

\begin{equation}
\Sigma _{\sigma \sigma ^{\prime }}^{\mathbf{k}}(\mathbf{r}_{\tau },\mathbf{r}%
_{\tau ^{\prime }}^{\prime },\omega )=\delta _{\tau t}\delta _{\tau ^{\prime
}t^{\prime }}\sum_{aa^{\prime }}\phi _{at}(\mathbf{r}_{t})\Sigma _{a\sigma
ta^{\prime }\sigma ^{\prime }t^{\prime }}^{corr}(\mathbf{k},\omega )\phi
_{a^{\prime }t^{\prime }}^{\ast }(\mathbf{r}_{t^{\prime }}^{\prime }),
\label{EQ22}
\end{equation}%
and is only non--zero when the legs $\mathbf{r}$ $\mathbf{r}^{\prime }$ land
inside the correlated sites. A single--particle LDA\ Hamiltonian with
relativistic Pauli term is a 2x2 matrix operator 
\begin{equation}
\hat{H}_{\sigma \sigma ^{\prime }}=-\nabla ^{2}\delta _{\sigma \sigma
^{\prime }}+V_{\sigma \sigma ^{\prime }}^{LDA}(\mathbf{r}).  \label{EQ23}
\end{equation}%
Since LDA\ potential already includes correlations in some average form,
there exists a site diagonal double counting term $V_{a\sigma a^{\prime
}\sigma ^{\prime }}^{DC(t)}$ which has to be subtracted from the
self--energy $\Sigma _{a\sigma ta^{\prime }\sigma ^{\prime }t^{\prime
}}^{corr}(\mathbf{k},\omega )$ in Eq.(\ref{EQ22}). There is a vast
literature about it, therefore here we ignore this subject and refer the
reader to a recent work and references therein \cite{KristjanDC}.

We represent the Green function of the lattice in terms of some, possibly
non--orthonormal, basis set $\chi _{\alpha \tau }^{\mathbf{k}}(\mathbf{r}),$
such as the one used in a full potential multiple--$\kappa $ linear
muffin--tin orbital (LMTO) method \cite{FPLMTO}, as follows 
\begin{equation}
G_{\sigma \sigma ^{\prime }}^{\mathbf{k}}(\mathbf{r},\mathbf{r}^{\prime
},\omega )=\sum_{\alpha \tau \alpha ^{\prime }\tau ^{\prime }}\chi _{\alpha
\tau }^{\mathbf{k}}(\mathbf{r})G_{\alpha \sigma \tau \alpha ^{\prime }\sigma
^{\prime }\tau ^{\prime }}(\mathbf{k},\omega )\chi _{\alpha ^{\prime }\tau
^{\prime }}^{\mathbf{k\ast }}(\mathbf{r}^{\prime }).  \label{EQ24}
\end{equation}%
The inverse of the interacting Green function is the matrix 
\begin{eqnarray}
G_{\alpha \sigma \tau \alpha ^{\prime }\sigma ^{\prime }\tau ^{\prime
}}^{-1}(\mathbf{k},\omega ) &=&\langle \chi _{\alpha \tau }^{\mathbf{k}%
}|(\omega +\epsilon _{F})\delta _{\sigma \sigma ^{\prime }}-\hat{H}_{\sigma
\sigma ^{\prime }}-\hat{\Sigma}_{\sigma \sigma ^{\prime }}+\hat{V}_{\sigma
\sigma ^{\prime }}^{DC}|\chi _{\alpha ^{\prime }\tau ^{\prime }}^{\mathbf{k}%
}\rangle  \notag \\
&=&(\omega +\epsilon _{F})O_{\alpha \sigma \tau \alpha ^{\prime }\sigma
^{\prime }\tau ^{\prime }}^{\mathbf{k}}-H_{\alpha \sigma \tau \alpha
^{\prime }\sigma ^{\prime }\tau ^{\prime }}^{\mathbf{k}}-\Sigma _{\alpha
\sigma \tau \alpha ^{\prime }\sigma ^{\prime }\tau ^{\prime }}^{\mathbf{k}%
}(\omega )+V_{\alpha \sigma \tau \alpha ^{\prime }\sigma ^{\prime }\tau
^{\prime }}^{\mathbf{k,}DC}.  \label{EQ25}
\end{eqnarray}%
It is expressed via the matrix elements of the LDA\ Hamiltonian, the overlap
matrix, the correlated block of the self--energy $\Sigma _{a\sigma
ta^{\prime }\sigma ^{\prime }t^{\prime }}^{corr}(\mathbf{k},\omega )$ and of
the double counting potential, as follows

\begin{eqnarray}
H_{\alpha \sigma \tau \alpha ^{\prime }\sigma ^{\prime }\tau ^{\prime }}^{%
\mathbf{k}} &=&\langle \chi _{\alpha \tau }^{\mathbf{k}}|\hat{H}_{\sigma
\sigma ^{\prime }}|\chi _{\alpha ^{\prime }\tau ^{\prime }}^{\mathbf{k}%
}\rangle ,  \label{EQ261} \\
O_{\alpha \sigma \tau \alpha ^{\prime }\sigma ^{\prime }\tau ^{\prime }}^{%
\mathbf{k}} &=&\delta _{\sigma \sigma ^{\prime }}\langle \chi _{\alpha \tau
}^{\mathbf{k}}|\chi _{\alpha ^{\prime }\tau ^{\prime }}^{\mathbf{k}}\rangle ,
\label{EQ262} \\
\Sigma _{\alpha \sigma \tau \alpha ^{\prime }\sigma ^{\prime }\tau ^{\prime
}}^{\mathbf{k}}(\omega ) &=&\sum_{ata^{\prime }t^{\prime }}\langle \chi
_{\alpha \tau }^{\mathbf{k}}|\phi _{at}\rangle \Sigma _{a\sigma ta^{\prime
}\sigma ^{\prime }t^{\prime }}^{corr}(\mathbf{k},\omega )\langle \phi
_{a^{\prime }t^{\prime }}|\chi _{\alpha ^{\prime }\tau ^{\prime }}^{\mathbf{k%
}}\rangle ,  \label{EQ263} \\
V_{\alpha \sigma \tau \alpha ^{\prime }\sigma ^{\prime }\tau ^{\prime }}^{%
\mathbf{k,}DC} &=&\sum_{aa^{\prime }}\langle \chi _{\alpha \tau }^{\mathbf{k}%
}|\phi _{at}\rangle V_{a\sigma a^{\prime }\sigma ^{\prime }}^{DC(t)}\langle
\phi _{a^{\prime }t}|\chi _{\alpha ^{\prime }\tau ^{\prime }}^{\mathbf{k}%
}\rangle .  \label{EQ264}
\end{eqnarray}

We note that the k--dependence of the matrix element $\Sigma _{\alpha \sigma
\tau \alpha ^{\prime }\sigma ^{\prime }\tau ^{\prime }}^{\mathbf{k}}(\omega
) $ comes here from both the non--trivial behavior for $\Sigma _{a\sigma
ta^{\prime }\sigma ^{\prime }t^{\prime }}^{corr}(\mathbf{k},\omega )$ as
well as from the projector $\langle \chi _{\alpha }^{\mathbf{k}}|\phi
_{at}\rangle .$ Therefore, even for methodologies utilizing the local
approximations, such as LDA+U and LDA+DMFT, the corresponding poles of the
single--particle Green functions acquire the k--dependence induced by the
hybridization effects with non--interacting electrons.

Given the prescription for computing the matrix $\Sigma _{a\sigma ta^{\prime
}\sigma ^{\prime }t^{\prime }}^{corr}(\mathbf{k},\omega )$, the poles of the
single--particle Green function can, for example, be analyzed by
diagonalizing the non--hermitian matrix $H_{\alpha \sigma \tau \alpha
^{\prime }\sigma ^{\prime }\tau ^{\prime }}^{\mathbf{k}}+\Sigma _{\alpha
\sigma \tau \alpha ^{\prime }\sigma ^{\prime }\tau ^{\prime }}^{\mathbf{k}%
}(\omega )-V_{\alpha \sigma \tau \alpha ^{\prime }\sigma ^{\prime }\tau
^{\prime }}^{\mathbf{k,}DC}$ for each frequency $\omega $ with the help of
its right and left eigenvectors%
\begin{eqnarray}
\sum_{\alpha ^{\prime }\sigma ^{\prime }\tau ^{\prime }}[H_{\alpha \sigma
\tau \alpha ^{\prime }\sigma ^{\prime }\tau ^{\prime }}^{\mathbf{k}}+\Sigma
_{\alpha \sigma \tau \alpha ^{\prime }\sigma ^{\prime }\tau ^{\prime }}^{%
\mathbf{k}}(\omega )-V_{\alpha \sigma \tau \alpha ^{\prime }\sigma ^{\prime
}\tau ^{\prime }}^{\mathbf{k,}DC}-p_{\mathbf{k}j}(\omega )O_{\alpha \sigma
\tau \alpha ^{\prime }\sigma ^{\prime }\tau ^{\prime }}^{\mathbf{k}%
}]R_{\alpha ^{\prime }\sigma ^{\prime }\tau ^{\prime }}^{\mathbf{k}j}(\omega
) &=&0,  \label{EQ271} \\
\sum_{\alpha \sigma \tau }L_{\alpha \sigma \tau }^{\mathbf{k}j}(\omega
)[H_{\alpha \sigma \tau \alpha ^{\prime }\sigma ^{\prime }\tau ^{\prime }}^{%
\mathbf{k}}+\Sigma _{\alpha \sigma \tau \alpha ^{\prime }\sigma ^{\prime
}\tau ^{\prime }}^{\mathbf{k}}(\omega )-V_{\alpha \sigma \tau \alpha
^{\prime }\sigma ^{\prime }\tau ^{\prime }}^{\mathbf{k,}DC}-p_{\mathbf{k}%
j}(\omega )O_{\alpha \sigma \tau \alpha ^{\prime }\sigma ^{\prime }\tau
^{\prime }}^{\mathbf{k}}] &=&0.  \label{EQ272}
\end{eqnarray}

\subsection{FLEX Self--Energy}

The prescription for computing the matrix $\Sigma _{a\sigma ta^{\prime
}\sigma ^{\prime }t^{\prime }}^{corr}(\mathbf{k},\omega )$ within the subset
of correlated electrons can be obtained by a variety of methods. The
dynamical mean field theory uses a k--independent approximation: $\Sigma
_{a\sigma ta^{\prime }\sigma ^{\prime }t^{\prime }}^{corr}(\mathbf{k},\omega
)\equiv \delta _{tt^{\prime }}\Sigma _{a\sigma a^{\prime }\sigma ^{\prime
}}^{DMFT(t)}(\omega )$ and solves the corresponding Anderson impurity
problem subjected to a self--consistency condition$.$ The treatment of the
substitutional site disorder can utilize a coherent potential approximation
(CPA)\cite{CPA}, $\Sigma _{a\sigma ta^{\prime }\sigma ^{\prime }t^{\prime
}}^{corr}(\mathbf{k},\omega )\equiv \delta _{tt^{\prime }}\Sigma _{a\sigma
a^{\prime }\sigma ^{\prime }}^{CPA(t)}(\mathbf{k},\omega ),$ where the
subset \{$a$\} should, in principle, refer to all orbitals (not only the
ones restricted by a particular angular momentum $l$) within the substituted
site $t$ of the lattice. The technique is similar to DMFT as it has been
recently implemented for studies of surface vacancies in TaAs Weyl semimetal 
\cite{TaAs}. The fluctuational exchange approximation relies on the
diagrammatic summation of the bubble and ladder diagrams: $\Sigma _{a\sigma
ta^{\prime }\sigma ^{\prime }t^{\prime }}^{corr}(\mathbf{k},\omega )\equiv
\Sigma _{a\sigma ta^{\prime }\sigma ^{\prime }t^{\prime }}^{FLEX}(\mathbf{k}%
,\omega )$.

We now briefly describe our implementation for computing the FLEX
self--energy. All calculations are done at real frequency axis at $T=0.$We
neglect by the particle--particle ladders which are known to be small, at
least for the problem of paramagnons where the most divergent term is given
by the particle--hole ladders. Contrary to the bubble diagrams which are
expressed via two--point functions in the real space, the ladder diagrams\
rely on the 4--point functions in general, but the use of the on--site
Hubbard--type interactions allows one to express all quantities via the
charge and spin (longitudinal and transverse) susceptibilities which are the
two--point functions. Despite this simplification gives the scaling with the
number of atoms in the unit cell as $N_{\{\tau \}}^{2},$ it is still a
computationally very demanding problem because the number of matrix elements
for representing the susceptibility grows as $N_{\{\tau \}}^{2}N_{orb}^{4}$
where $N_{orb}$ is the size of complete orbital manifold per atom needed.
This, for example, slows down the calculation with the GW method. However,
the restriction by the correlated subset simplifies the calculation
tremendously, because now the susceptibility matrices have the size $%
N_{\{t\}}^{2}N_{corr}^{4}.$

We now define the susceptibility within the correlated subset. It is
represented by the convolution of two Green functions on the frequency axis 
\begin{equation}
\pi _{a\sigma bst,b^{\prime }s^{\prime }a^{\prime }\sigma ^{\prime
}t^{\prime }}(\mathbf{q},\omega )=-i\sum_{\mathbf{k}}\int_{-\infty
}^{+\infty }\frac{d\omega ^{\prime }}{2\pi }G_{bstb^{\prime }s^{\prime
}t^{\prime }}(\mathbf{k+q},\omega ^{\prime })G_{a^{\prime }\sigma ^{\prime
}t^{\prime }a\sigma t}(\mathbf{k},\omega +\omega ^{\prime })e^{i\omega
^{\prime }0^{+}}.  \label{EQ361}
\end{equation}%
For the non--interacting (LDA) Green's functions%
\begin{equation}
G_{a^{\prime }\sigma ^{\prime }t^{\prime }a\sigma t}(\mathbf{k},\omega
)\rightarrow G_{a^{\prime }\sigma ^{\prime }t^{\prime }a\sigma t}^{(0)}(%
\mathbf{k},\omega )=\sum_{j}\frac{\langle \phi _{a^{\prime }t^{\prime
}}|\psi _{\mathbf{k}j\sigma ^{\prime }}\rangle \langle \psi _{\mathbf{k}%
ja}|\phi _{\sigma t}\rangle }{\omega -\epsilon _{\mathbf{k}j}-i0^{+}\mathrm{%
sign}(\epsilon _{F}-\epsilon _{\mathbf{k}j})},  \label{EQ37}
\end{equation}%
represented in the basis of the Bloch wave functions that diagonalize the
LDA\ Hamiltonian%
\begin{eqnarray}
\psi _{\mathbf{k}j\sigma }(\mathbf{r}) &=&\sum_{\alpha \tau }A_{\alpha
\sigma \tau }^{\mathbf{k}j}\chi _{\alpha \tau }^{\mathbf{k}}(\mathbf{r}),
\label{EQ291} \\
0 &=&\sum_{\alpha ^{\prime }\sigma ^{\prime }\tau ^{\prime }}(H_{\alpha
\sigma \tau \alpha ^{\prime }\sigma ^{\prime }\tau ^{\prime }}^{\mathbf{k}%
}-\epsilon _{\mathbf{k}j}O_{\alpha \tau \alpha ^{\prime }\tau ^{\prime }}^{%
\mathbf{k}})A_{\alpha ^{\prime }\sigma ^{\prime }\tau ^{\prime }}^{\mathbf{k}%
j},  \label{EQ292}
\end{eqnarray}%
the resulting expression for susceptibility matrix elements is given by%
\begin{eqnarray}
\pi _{a\sigma bst,b^{\prime }s^{\prime }a^{\prime }\sigma ^{\prime
}t^{\prime }}(\mathbf{q},\omega ) &=&\sum_{\mathbf{k}jj^{\prime }}\frac{f_{%
\mathbf{k}j}-f_{\mathbf{k}+\mathbf{q}j^{\prime }}}{\omega +\epsilon _{%
\mathbf{k}j}-\epsilon _{\mathbf{k+q}j^{\prime }}+i0^{+}\mathrm{sign}%
(\epsilon _{F}-\epsilon _{\mathbf{k}j})-i0^{+}\mathrm{sign}(\epsilon
_{F}-\epsilon _{\mathbf{k}+\mathbf{q}j^{\prime }})}\times  \notag \\
&&\langle \psi _{\mathbf{k}j\sigma }|\phi _{at}\rangle \langle \phi
_{bt}|\psi _{\mathbf{k+q}j^{\prime }s}\rangle \langle \psi _{\mathbf{k+q}%
j^{\prime }s^{\prime }}|\phi _{b^{\prime }t^{\prime }}\rangle \langle \phi
_{a^{\prime }t^{\prime }}|\psi _{\mathbf{k}j\sigma ^{\prime }}\rangle .
\label{EQ30}
\end{eqnarray}%
We note that exactly as in the spirit of the LDA+U and LDA+DMFT methods, the
index $a\sigma bs$ here describes the active Hilbert space of the atom $t$,
where $a$ and $b$ are orbital while $\sigma $ and $s$ are spin degrees of
freedom. For $d$--electrons its size is only $(5\ast 2)^{2}=100.$ This is
much smaller of the full Hilbert space needed to describe the susceptibility
matrix.

We next introduce the on--site Hubbard matrix $U_{aba^{\prime }b^{\prime
}}^{(t)}$ which describes the Coulomb interaction matrix elements among
correlated orbitals%
\begin{equation}
\langle \phi _{a}\phi _{a^{\prime }}|\frac{e^{2}}{r}|\phi _{b}\phi
_{b^{\prime }}\rangle _{\Omega _{t}}=U_{aba^{\prime }b^{\prime }}^{(t)}.
\label{EQ31}
\end{equation}%
We use such parametrization so that the screening effects in $U$ can be
taken into account by an external calculation or phenomenologically. It is
useful to define the interaction as the difference between bare and exchange
terms and introduce spin indexes explicitly so that the indexation matches
the one for susceptibility%
\begin{equation}
I_{a\sigma bst,a^{\prime }\sigma ^{\prime }b^{\prime }s^{\prime }t^{\prime
}}=\delta _{tt^{\prime }}[\delta _{\sigma s}\delta _{\sigma ^{\prime
}s^{\prime }}U_{aba^{\prime }b^{\prime }}^{(t)}-\delta _{\sigma s^{\prime
}}\delta _{\sigma ^{\prime }s}U_{ab^{\prime }a^{\prime }b}^{(t)}]
\label{EQ32}
\end{equation}%
This allows to drop the indexation and manipulate with matrix products
symbolically. Define the dielectric function matrix for the correlated
subspace 
\begin{equation}
\hat{\epsilon}(\mathbf{q},\omega )=\hat{1}-\hat{I}\hat{\pi}(\mathbf{q}%
,\omega ).  \label{EQ33}
\end{equation}%
Its inverse gives rise to the interacting susceptibility%
\begin{equation}
\hat{\chi}(\mathbf{q},\omega )=\hat{\pi}(\mathbf{q},\omega )\hat{\epsilon}%
^{-1}(\mathbf{q},\omega ),  \label{EQ34}
\end{equation}%
and to the dynamically screened interaction for the correlated manifold%
\begin{equation}
\hat{K}(\mathbf{q},\omega )=\hat{I}+\hat{I}[\hat{\chi}(\mathbf{q},\omega )-%
\frac{1}{2}\hat{\pi}(\mathbf{q},\omega )]\hat{I}.  \label{EQ35}
\end{equation}%
The subtraction of $\frac{1}{2}\hat{\pi}(\mathbf{q},\omega )$ takes care of
the single bubble diagram that appears twice in both bubble and ladder
series.

The FLEX\ self--energy appears as the integral over frequencies

\begin{equation}
\Sigma _{a\sigma ta^{\prime }\sigma ^{\prime }t^{\prime }}^{FLEX}(\mathbf{k}%
,\omega )=-\sum_{bb^{\prime }}\sum_{ss^{\prime }}\sum_{\mathbf{q}%
}\int_{-\infty }^{+\infty }\frac{d\omega ^{\prime }}{2\pi i}K_{a\sigma
bst,b^{\prime }s^{\prime }a^{\prime }\sigma ^{\prime }t^{\prime }}(\mathbf{k}%
-\mathbf{q},\omega ^{\prime })G_{bstb^{\prime }s^{\prime }t^{\prime }}^{(0)}(%
\mathbf{q},\omega +\omega ^{\prime })e^{i\omega ^{\prime }0^{+}}.
\label{EQ36}
\end{equation}%
Here we have used the non--interacting LDA Green function $G^{(0)}$ within
the correlated subset, Eq. (\ref{EQ37}). To evaluate the frequency integral
in practice, we use spectral representation for the dynamically screened
interaction $K\ $which allows to perform integration over frequencies
analytically. This is similar how it is sometimes done in GW\ implementations%
\cite{GW}.

Finally, we check the Hartree--Fock limit and show that the FLEX\
self--energy goes exactly to the one used in the LDA+U method. Replace the
interaction matrix by site diagonal, frequency and $\mathbf{q}$ independent
matrix $I$ in Eq.(\ref{EQ36}) 
\begin{equation}
K_{a\sigma bst,a^{\prime }\sigma ^{\prime }b^{\prime }s^{\prime }t^{\prime
}}(\mathbf{q},\omega )\rightarrow I_{a\sigma bst,a^{\prime }\sigma ^{\prime
}b^{\prime }s^{\prime }t^{\prime }}\equiv \delta _{tt^{\prime }}I_{a\sigma
bs,a^{\prime }\sigma ^{\prime }b^{\prime }s^{\prime }}^{(t)}.  \label{EQ38}
\end{equation}%
The frequency integral in Eq. (\ref{EQ36}) is performed by closing the
contour in the upper plane due to $e^{i\omega ^{\prime }0^{+}}$. Then, the
only poles in the upper plane, (\textit{i.e.} those corresponding to the
occupied states), contribute and we obtain the definition of the density
matrix for correlated electrons%
\begin{eqnarray}
\sum_{\mathbf{k}}\int_{-\infty }^{+\infty }\frac{d\omega ^{\prime }}{2\pi i}%
G_{bstb^{\prime }s^{\prime }t^{\prime }}(\mathbf{k},\epsilon )e^{i\omega
^{\prime }0^{+}} &=&\sum_{\mathbf{k}j}\langle \phi _{bt}|\psi _{\mathbf{k}%
js}\rangle \langle \psi _{\mathbf{k}js^{\prime }}|\phi _{b^{\prime
}t^{\prime }}\rangle \int_{-\infty }^{+\infty }\frac{d\omega ^{\prime }}{%
2\pi i}\frac{e^{i\omega ^{\prime }0^{+}}}{\omega ^{\prime }-\epsilon _{%
\mathbf{k}j}-i0^{+}\mathrm{sign}(\epsilon _{F}-\epsilon _{\mathbf{k}j})} 
\notag \\
&=&\sum_{\mathbf{k}j}f_{\mathbf{k}j}\langle \phi _{bt}|\psi _{\mathbf{k}%
js}\rangle \langle \psi _{\mathbf{k}js^{\prime }}|\phi _{b^{\prime
}t^{\prime }}\rangle =n_{b^{\prime }s^{\prime }t^{\prime }bst}.
\label{EQ39X}
\end{eqnarray}%
The Hartree--Fock limit is now recovered%
\begin{eqnarray}
\Sigma _{a\sigma ta^{\prime }\sigma ^{\prime }t^{\prime }}^{FLEX}(\mathbf{k}%
,\omega ) &\rightarrow &\Sigma _{a\sigma ta^{\prime }\sigma ^{\prime
}t^{\prime }}^{LDA+U}=-\sum_{bb^{\prime }}\sum_{ss^{\prime }}I_{a\sigma
bst,b^{\prime }s^{\prime }a^{\prime }\sigma ^{\prime }t^{\prime
}}n_{b^{\prime }s^{\prime }t^{\prime }bst}  \notag \\
&=&\delta _{tt^{\prime }}\sum_{bb^{\prime }}\sum_{ss^{\prime }}I_{a\sigma
a^{\prime }\sigma ^{\prime },b^{\prime }s^{\prime }bs}^{(t)}n_{b^{\prime
}s^{\prime }bs}^{(t)},  \label{EQ40}
\end{eqnarray}%
where only site diagonal density matrix for the correlated electrons is
needed%
\begin{equation}
n_{b^{\prime }s^{\prime }bs}^{(t)}=n_{b^{\prime }s^{\prime }tbst}.
\label{EQ41}
\end{equation}%
This $\Sigma ^{LDA+U}$ is used in the LDA+U\ method.

\end{widetext}

\subsection{Note on Self--Consistency}

We now comment on the self--consistency condition within this approach.
First, due to the existence of generating functionals for both GW and FLEX
approximations\cite{GW,FLEX}, it looks like the self--consistency with
respect to the Green functions and the interactions has to be implemented.
However, at least within the GW method the subject was studied in some
details with applications to some real materials\cite{SCF-GW}. The short
answer is that finding fully self--consistent solution within the bubble
diagrams is not a good idea because while providing better total energies,
it worsens the one--electron spectra. There is also a technical part of the
problem that once the complex self--energy is introduced, the polarizability
(\ref{EQ361}) can no longer be represented in its simple form (\ref{EQ30})
and alternative formulations via, for example, imaginary time axis need to
be implemented.

The self--consistency is however an important step within DMFT as it allows
to describe, for example, the Mott transition. One can easily combine the
non--local FLEX self--energy with the DMFT local self--energy, in accord
with the recent proposals \cite{FLEX+DMFT,Danger} 
\begin{eqnarray}
\Sigma ^{DMFT+FLEX}(\mathbf{k},\omega ) &=&\Sigma ^{DMFT}(\omega )+  \notag
\\
&&\Sigma ^{FLEX}(\mathbf{k},\omega )-\Sigma ^{DC}(\omega ).  \label{EQ42}
\end{eqnarray}%
This\ allows to utilize sophisticated impurity solvers developed in DMFT
community. Here, the subtracted double counting term $\Sigma ^{DC}(\omega )$
utilizes the FLEX approximation itself as the impurity solver \cite%
{FLEX+DMFT}, where one calculates the local polarizability $\pi
_{loc}(\omega )=\sum_{q}\pi (\mathbf{q},\omega )$ which is represented in
this method as the product of the two local Green functions, that will
subsequently appear in Eq. (\ref{EQ361}) once the summation over $\mathbf{q}$
is performed. Then, the local interaction, as in Eq. (\ref{EQ35}), is
computed which gives rise to the local impurity self--energy within the FLEX
approximation. We denote it hereafter as $\Sigma ^{DC}(\omega )=\Sigma
^{DMFT(FLEX)}(\omega ).$ Another option for $\Sigma ^{DC}(\omega )$ is to
use the local FLEX self--energy \cite{Danger} 
\begin{equation}
\Sigma _{loc}^{FLEX}(\omega )=\sum_{\mathbf{k}}\Sigma ^{FLEX}(\mathbf{k}%
,\omega ).  \label{EQ43}
\end{equation}%
Note that DMFT(FLEX) approach has been recently applied to study the mass
enhancement in Pd \cite{PdDMFT}. We discuss the outcomes of various
approximations for the self--energy in V and Pd in the following section.

Another sort of self--consistency that was developed in the past is the
quasiparticle self--consistency. That is when not the full self--energy but
its value at $\omega =0$ and its frequency derivative around $\omega =0$
describing mass enhancement data are used to reconstruct new densities and
resulting quasiparticle Green's functions. It was developed in connection
with the GW\ approach, and was shown to reproduce the band gaps of
semiconductors quite well \cite{QSGW}. A combination of the LDA and
Gutzwiller's method (called LDA+G) explores a similar idea \cite{LDA+G}
where the variational Gutzwiller method is used to find those self--energy
parameters. It was also implemented in a most recent combination of the GW
and DMFT called QSGW+DMFT\cite{QSGW+DMFT}.

It is fairly straightforward to implement this sort of self--consistency
within the described LDA+FLEX method. The polarizability is still
represented by its non--interacting form (\ref{EQ30}) since quasiparticle
approximation for the Green functions assumes real eigenvalues. It is also
easy to update the position of the Fermi level and restore the new density
at each iteration which replaces the LDA\ density in Eq. (\ref{EQ23}).
However, in our applications to V and Pd we did not find any noticeable
changes in the obtained self--energies and the spectral functions for
d--electrons when doing these updates, and the results obtained at first
iteration by using the LDA\ band structures are already very close
self--consistency. It would probably make more sense doing it when
evaluating total energies but this topic is beyond the scope of the present
work.

\section{III. Applications to V and Pd}

Here we describe applications of our LDA+FLEX implementation to V and Pd.
These two elemental metals have been at the center of interest for a long
time, and, in particular Palladium, whose strong spin fluctuations are known
to destroy superconductivity \cite{BerkSchrieffer} and contribute to
specific heat renormalization by $\lambda \sim 0.3-0.4$\cite%
{BerkSchrieffer,Winter,Pinski,Savrasov}. Most recent LDA+DMFT study\cite%
{PdDMFT} addressed the mass enhancement of Pd in great detail but extracted
smaller $\lambda =0.03-0.09$ for the values of $U$ ranging from 1 to 4 eV
using LDA+DMFT method with the FLEX approximation as the impurity solver.
Vanadium is known to be less paramagnetically enhanced and its specific heat
renormalization may well be described electron--phonon interactions alone%
\cite{Savrasov}. However, a room for spin--fluctuational contribution still
exists as $\lambda $ based on the FLEX\ calculated self--energy with
Stoner--type LDA interaction strenght was earlier found to be 0.2 \cite%
{Winter}. It is also known that one needs a pretty large value of the
effective Coulomb pseudopotential $\mu ^{\ast }\sim 0.3$ to adjust the
superconducting critical temperature of V to the one known from experiment 
\cite{Savrasov}, part of which may be attributed to $\lambda .$

\begin{figure*}[tbh]
\includegraphics[height=0.59\textwidth,width=0.80\textwidth]{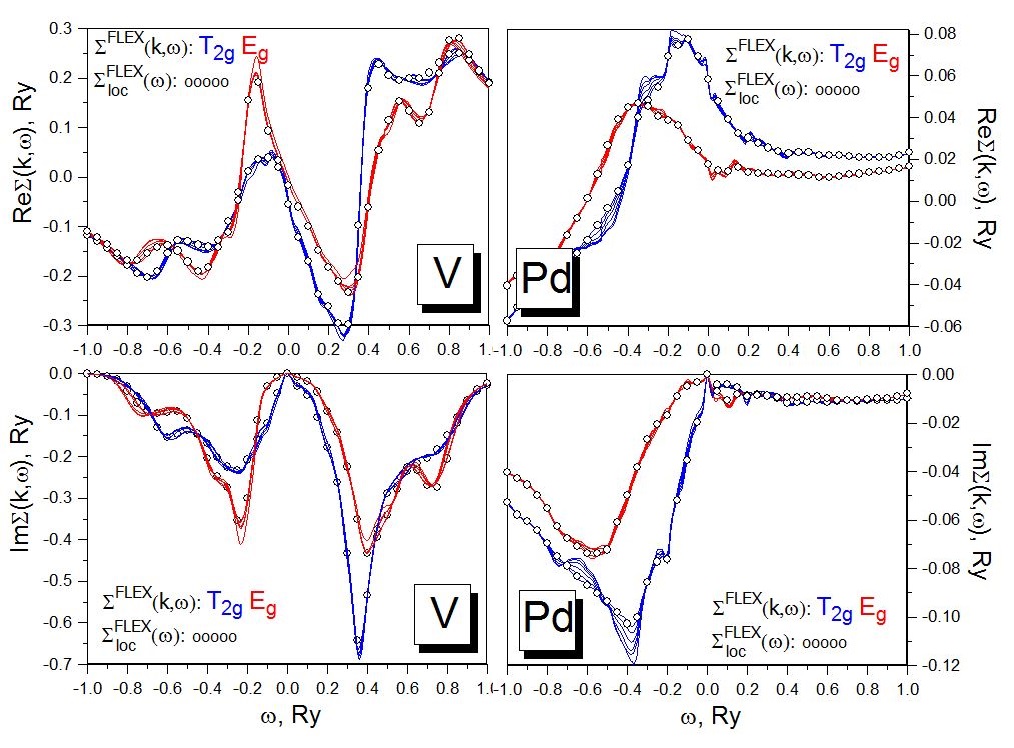}
\caption{Calculated self--energy $\Sigma (\mathbf{k},\protect\omega )$ (top
is the real part, and bottom is imaginary part) using FLEX approximation for
t$_{2g}$ and e$_{g}$ electrons in V and Pd for the wavevector k traversing
along (0$\protect\xi \protect\xi $) direction of the Brillouin Zone. The
circles show the result of the sum over all k--points representing the local
FLEX self--energy. A representative value of Hubbard $U$=0.2 Ry is used in
these plots, but the notion of locality persists for a whole range of U's.}
\label{FigFLEX}
\end{figure*}

For our band structure calculational framework we use double--$\kappa $ full
potential LMTO\ method as implemented in Ref. \cite{FPLMTO}. The Green
functions, susceptibilities and interactions are computed on the grid of 400
frequencies and for 256 non--equivalent wave vectors set by (20,20,20)
divisions of the reciprocal unit cell. All integrals over the BZ are
performed using grids set by (60,60,60) divisions of the reciprocal unit
cell with help of a version of the tetrahedron method adapted for linear
response calculations \cite{FPLMTO}.

As far as determining precise value of $U$ for these materials, there is
some obvious uncertainty here. One estimate can be given by associating it
with the Stoner parameter which was calculated for these metals to be as
small as 0.025 Ry (0.34 eV) \cite{Janak}. The upper estimate can also be
obtained from the Stoner criterion of magnetism, i.e. when the static
interacting susceptibility as given by Eq. (\ref{EQ34}) diverges. We have
analyzed eigenvalues of the wavevector dependent dielectric matrix, Eq. (\ref%
{EQ33}), at $\omega =0$ in the active space of $d$--electrons and found that
the negative eigenvalues appear at $U_{c}=0.31$\ Ry for V and at $0.26$ Ry
for Pd. These critical values should signalize that the system undergoes the
spin density wave (SDW) transition within this approach. (The use of local
quantities in Stoner criterion, \textit{i.e.} the ones summed over $\mathbf{q%
}$, lead to $U_{c}=0.43$\ Ry for V and at $0.59$ Ry for Pd). We perform all
computations for a range of $U$ values varying it from 0 to 0.2 Ry.

We now discuss our calculated d--electron self--energies. Our results for V
and Pd are shown in Fig. \ref{FigFLEX} where we plot matrix elements of 
\textrm{Re}$\Sigma ^{FLEX}(\mathbf{k},\omega )$ and \textrm{Im}$\Sigma
^{FLEX}(\mathbf{k},\omega )$ for both T$_{2g}$ and E$_{g}$ electrons. Here
we use some representative value of $U$=0.2 Ry (2.7 eV) but our conclusions
remain the same for the whole range of $U$'s that we studied. The Hartree
Fock values for $\text{Re}\Sigma $ are subtracted in both plots. To
illustrate the k--dependence, the self--energies are given for several wave
vectors k chosen along $(0\xi \xi )$ direction of the Brillouin Zone (BZ).
It is remarkable that the k--dependence in these plots is seen to be quite
small prompting that the local self--energy approximation may be adequate.
We also compared the self--energies for other k--points in the BZ and came
up with a similar conclusion. We subsequently calculate the $\Sigma
_{loc}^{FLEX}(\omega )$ as a sum over k--points whose frequency dependence
is also visualized in Fig. \ref{FigFLEX} by small circles. We see a close
agreement between $\Sigma _{loc}^{FLEX}(\omega )$ and $\Sigma ^{FLEX}(%
\mathbf{k},\omega )$ for both T$_{2g}$ and E$_{g}$ matrix elements.

\begin{figure}[tbp]
\includegraphics[height=0.64\textwidth,width=0.40\textwidth]{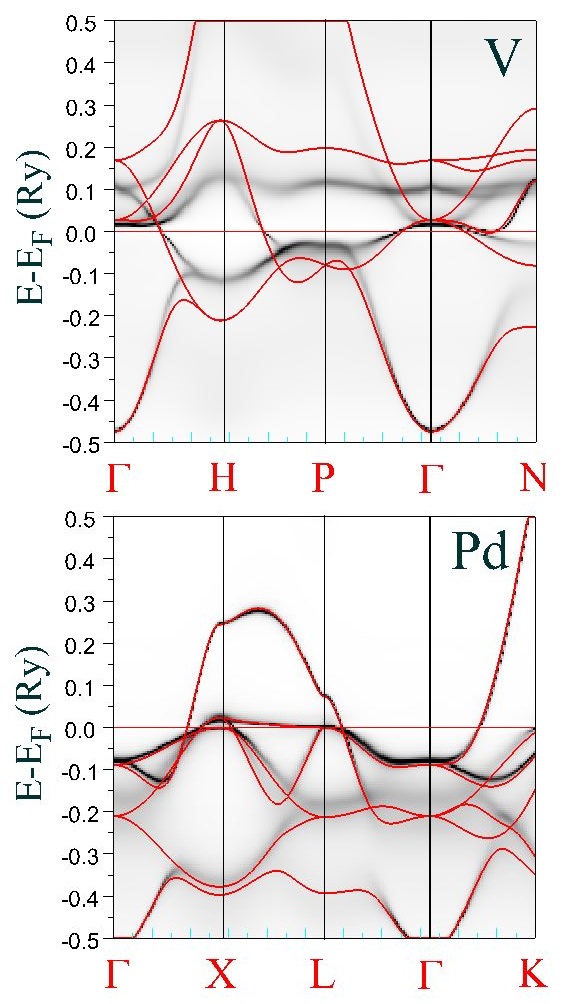}
\caption{Effect of FLEX self-energy on the calculated poles of single
particle Green's functions for V and Pd as compared with their LDA\ band
structures . The local FLEX value at $\protect\omega =0$ is subtracted from $%
\Sigma ^{FLEX}(\mathbf{k},\protect\omega )$ when visualizing $\text{Im}G(%
\mathbf{k},\protect\omega ).$ Hubbard $U$=0.2 Ry is used.}
\label{FigBands}
\end{figure}

Based on our calculated d--electron self--energies $\Sigma ^{FLEX}(\mathbf{k}%
,\omega )$, we calculate poles of single particle Greens functions and plot
the obtained $\text{Im}G(\mathbf{k},\omega )$ for V and Pd in Fig. \ref%
{FigBands}. Here we compare the results of our many--body calculation with
the energy band structures obtained by LDA. Although many versions of the
double counting potentials currently exist\cite{KristjanDC}, to illustrate
the $\mathbf{k-}$ and $\omega $ dependence of the FLEX self--energy we
merely subtract from $\Sigma ^{FLEX}(\mathbf{k},\omega )$ its local value,
Eq. (\ref{EQ43}), at $\omega =0$ in order to visualize $\text{Im}G(\mathbf{k}%
,\omega )$. As one sees, the primary effect of the self--energy is the
renormalization of the d--electron bandwidth and a small broadening that is
acquired by the d--electrons due to the appearance of the imaginary part of $%
\Sigma ^{FLEX}(\mathbf{k},\omega )$. We use the same self--energies
(calculated at $U$=0.2 Ry) as plotted in Fig. \ref{FigFLEX}.

We further would like to compare the results of our calculations with the
self--energies obtained using DMFT and directly with experiment.
Unfortunately, most accurate solver developed to date, Continuous Time
Quantum Monte Carlo method\cite{CTQMC}, works on imaginary time--frequency
axis and obtaining frequency dependence of the self--energy on the real axis
involves an analytical continuation algorithm which is known to be not very
accurate. However, one can easily perform calculations of correlation
induced electronic mass enhancement in both metals using DMFT(QMC) since it
can be extracted directly from the knowledge of $\Sigma ^{DMFT(QMC)}(i\omega
_{n})$ on imaginary axis. The mass enhancement is then determined as the
Fermi surface average of the frequency derivative of the self--energy taken
at either $\omega =0$ or at $i\omega _{n}\rightarrow 0.$ This is also a more
sensitive way to compare various approximations for the self--energy. To
perform LDA+DMFT(QMC) calculations we downloaded Embedded DMFT code
developed by Haule \cite{Haule}. 
\begin{figure}[tbp]
\includegraphics[height=0.63\textwidth,width=0.4\textwidth]{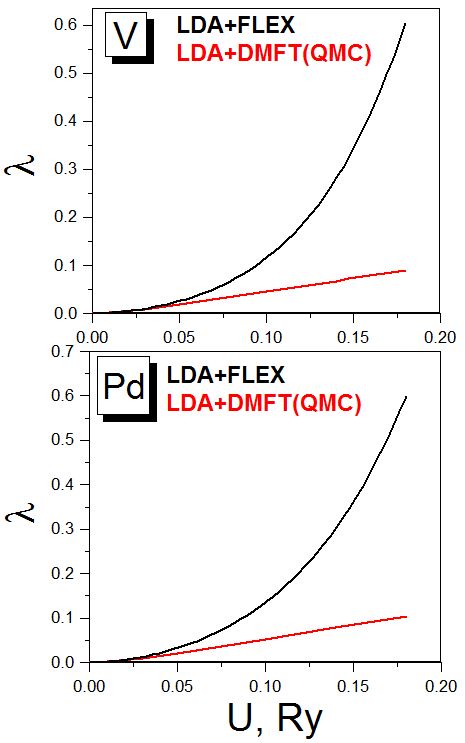}
\caption{Comparisons between LDA+FLEX and LDA+DMFT approximations for
predicting correlation induced electronic mass enhancement factor $\protect%
\lambda $ in V and Pd. The calculations with Dynamical Mean Field Theory are
performed using Quantum Monte Carlo method as implemented in Ref. 
\protect\cite{Haule}. }
\label{FigMass}
\end{figure}

Fig. \ref{FigMass} shows our calculated mass enhancements for V or Pd. To
gain some physical insights on approximations used in these simulations, we
choose to provide these data as a function of $U$. One can see that for
small values of $U$ both FLEX and DMFT give very similar mass enhancements.
This is quite easy to understand because when $U$ goes to zero, due to the
emergent locality of the self--energy evident from Fig. \ref{FigFLEX}, the
FLEX provides a good approximation for solving the impurity problem while
self--consistency imposed by DMFT is not essential \cite{RMP}. However, as $%
U $ increases, FLEX delivers significantly larger values of $\lambda $ than
DMFT$.$ This can also be understood since we start approaching the spin
density wave transition which occurs at $U_{c}=0.31$\ Ry for V and at $0.26$
Ry for Pd within FLEX.

A recent publication addressed the specific heat renormalization in Pd using
LDA+DMFT method with FLEX as the impurity solver. The deduced $\lambda $ was
found to be in the range $0.03-0.09$ for the values of $U$ between 1 to 4
eV. This is in accord with our LDA+DMFT(QMC) simulation as seen from Fig. %
\ref{FigMass}. At least, within LDA+DMFT(FLEX), the result for such small
mass enhancement can be interpreted from the Stoner criterion, as the use of
the local susceptibilities instead of momentum resolved ones in Eqs. (\ref%
{EQ33})--Eq. (\ref{EQ35}) pushes $U_{c}$ to $0.43$\ Ry for V and to $0.59$
Ry for Pd, so that in the neglect of the DMFT\ self--consistency we are
simply further away from the instability. Overall, the trends that we
monitor here are pretty much expectable from a vast amount of simulations
performed on models \cite{RMP}. Thus, one can conclude that from the
standpoint of the comparison with the experiment, both DMFT(QMC) and
DMFT(FLEX) calculations underestimate the mass enhancement of Pd, while our
full momentum resolved FLEX implementation is capable to provide a more
reliable estimate, at least for the range of the values of $U$ employed in
the present study. One can speculate that FLEX\ still gives an unreliable
result while DMFT needs a significantly larger value of $U$ to deal with
this problem but the possibility of magnetic ordering transition at larger $%
U $'s should not be overlooked. Another option is the need for
self--consistent treatment of spin fluctuations and electron--phonon
interactions while extracting the specific heat renormalization but this
study is well beyond the scope of the present work.

One can finally comment on the results of a recently proposed DMFT+FLEX
scheme \cite{FLEX+DMFT}, Eq. (\ref{EQ42}). As a result of the weak coupling
regime that we study here for V and Pd, we can approximate $\Sigma
^{DMFT}(\omega )$ by $\Sigma ^{DMFT(FLEX)}(\omega )=\Sigma ^{DC}(\omega )$,
and the mass enhancements that would be obtained within it will practically
coincide with the ones that we find within FLEX itself. If, on the other
hand, one uses \cite{Danger} $\Sigma ^{DC}(\omega )=\Sigma
_{loc}^{FLEX}(\omega )$, Eq. (\ref{EQ43}), and because the d--electron
self--energies $\Sigma ^{FLEX}(\mathbf{k},\omega )$ are found to be well
approximated by $\Sigma _{loc}^{FLEX}(\omega )$, the mass enhancements that
would be obtained now will be the ones that we find from DMFT, thus bringing
no advantage in such combination at least for the problem of Pd.

\section{IV. Conclusion}

In conclusion, by implementing a combination of $\mathbf{k}-$ and $\omega -$
dependent self--energy found from fluctuational exchange approximation with
LDA, we are able to incorporate the effect of momentum resolved spin
fluctuations on the calculated single particle spectra of real materials.
Applicability of the approach was demonstrated for two elemental metals, V
and Pd whose self--energies have been found remarkably k--independent
justifying the use of local approximations. However, we find corresponding
mass enhancement data to be different when comparing the results of our
calculations with local LDA+DMFT approach, where LDA+FLEX delivers better
agreement with experiment for the range of values of $U\lesssim 0.2$ $Ry$.
The method is naturally combinable with Dynamical Mean Field Theory and we
are hoping that such extension may provide additional clues on the
electronic properties of other classes of systems, such, \textit{e.g.}, as
unconventional superconductors, where one can track materials dependence of
the superconductivity, which is somewhat lacking when addressing this
problem using model Hamiltonians.

\section{Acknowledgement}

The authors are indebted to Motoharu Kitatani for discussing the details of
their DMFT+FLEX method. This work was supported by the US\ NSF DMR Grant No.
1411336. X. G. Wan was supported by NSF of China, Grant No. 11525417.

\end{document}